\documentclass[review]{elsarticle}
\usepackage{lineno,hyperref,amsmath,xcolor}
\journal{Journal of \LaTeX\ Templates}

\usepackage{xcolor}
\begin{document}
\begin{frontmatter}
\title{Influence of oceanic turbulence on propagation of Airy vortex beam carrying orbital angular momentum}
\author[address,address1]{Xinguang Wang}
\author[address1]{Zhen Yang}
\author[address,address1]{Shengmei Zhao \corref{mycorrespondingauthor}}
\cortext[mycorrespondingauthor]{Corresponding author}
\ead{zhaosm@njupt.edu.cn}
\address[address]{Institute of Signal Processing and Transmission, Nanjing University of Posts and Telecommunications(NUPT), Nanjing 210003, China}
\address[address1]{Key Lab of Broadband Wireless Communication and Sensor Network Technology of Ministry of Education, NUPT, Nanjing, 210003, China}
\begin{abstract}
With Rytov approximation theory, we derive the analytic expression of detection probability of Airy vortex beam carrying orbital angular momentum (OAM) through an anisotropic weak oceanic turbulence. We investigate the influences of turbulence parameters and beam parameters on the propagation properties of Airy-OAM beam. The numerical simulation results show that the anisotropic oceanic turbulence with a lower dissipation rate of temperature variance, smaller ratio of temperature and salinity contributions to the refractive index spectrum, higher dissipation rate of kinetic energy per unit mass of fluid, bigger inner scale factor, larger anisotropic coefficient causes the larger detection probability of Airy-OAM beam. Moreover, the Airy-OAM beam with a smaller topological charge, larger main ring radius and longer wavelength, has strong resistance to oceanic turbulent interference. Additionally, the detection probability decreases with the increase of receiving aperture size. In comparison with Laguerre-Gaussian-OAM beam, Airy-OAM beam has more anti-interference to turbulence when its topological charge is larger than 5 due to its non-diffraction and self-healing characteristics. The results are useful for underwater optical communication link using Airy-OAM beam.
\end{abstract}

\end{frontmatter}

\section{Introduction}
With the growing demand of underwater optical communication(UOC), as well as the increasing needs of underwater imaging systems and sensor networks, the propagation properties of vortex beams carrying orbital angular momentum (OAM) has attracted a wider attention in an underwater environment
 \cite{14huangypoe,14xujiaolt,15liudajunao,15josaa,16chengmingjianao,16liudajunoptik,16turbid,chengmingjianIEEE,lorentz,hermite,zhaodaomu,sibangaosi,lommel,hankel,yangtianxing, pansunxiang}. The OAM modes with different topological charges are orthogonal to each other, which make it possible for OAM modes used as a new degree of freedom for information multiplexing, where the capacity, as well as the bandwidth efficiency, can be greatly enhanced.  Baghdady \emph{et al.} realized 3Gbit/s UOC system with two OAM modes multiplexing for link distance $2.96m$ by taking advantage of this characteristics of OAM modes \cite{bag}. Similarly, Ren \emph{et al.} achieved a higher transmission rate, 40Gbit/s, for a UOC system by using four OAM multiplexed modes \cite{sr}.


However, OAM mode is a spatial distribution, its wavefront is susceptible to the spatial aberrations caused by underwater or atmosphere turbulence\cite{1zu,2zu,3zu,4zu,5zuwai}.
That is, as optical signals carrying OAM propagation through an oceanic medium,
it will suffer attenuation and wavefront distortion caused by the fluctuations of the refractive
index of water and the various constituents in the ocean \cite{2000,zhaols}, which results in the OAM
crosstalk between modes and diminish the performance of an optical communication system. 
Importantly, the existed results show that different OAM beams have different propagation properties in the underwater environment. For example, Huang \emph{et al.} investigated the propagation of Gaussian Schell-model vortex beams through oceanic turbulence, and showed that both position and number of coherent vortices were changed with the increasing of propagation distance \cite{14huangypoe}. The propagation of a partially coherent cylindrical vector Laguerre-Gaussian(LG) beam passing through oceanic turbulence was investigated in \cite{15josaa}, and the results showed that the smaller the initial coherence length of beam was, the larger the influence of ocean turbulence was. Cheng \emph{et al.} revealed that the effect of a partially coherent LG beam with longer wavelength and smaller topological charge was less affected by ocean turbulence \cite{16chengmingjianao}, Bessel-Gaussian(BG) beam is better than LG beam to resist the effects of ocean turbulence due to the non-diffraction and self-healing properties \cite{chengmingjianIEEE}. As for partially coherent Lorentz-Gauss OAM beam, Liu \emph{et al. }found that the effect of turbulence was greater with smaller topological charge of beam\cite{lorentz}.

On the other hand, Airy vortex beam carrying OAM mode, called Airy-OAM beam, has the properties of non-diffraction, self-healing and self-accelerating. It has attracted a lot of attentions on its generation, properties and potential applications recently \cite{1airyjieshao,2airyjieshao,3airyjieshao}. However, as far as we have known, the propagation properties of Airy-OAM beam in underwater turbulence have not been reported.


In this paper, we investigate the propagation properties of Airy-OAM beam through anisotropic oceanic turbulence. The detection probability of Airy-OAM beam at the receiver side is derived with Rytov approximation theory, and the influence of oceanic turbulence on Airy-OAM beam with different oceanic environment and different source parameters are presented by numerical simulations.

The organization of the paper is as follows. In Section 2, the detection probability of Airy-OAM mode in weak oceanic turbulence is analyzed. In Section 3, the performance of Airy-OAM beam propagating in oceanic turbulence is discussed. Finally, Section 4 concludes the paper.

\section{The detection probability of Airy-OAM beam in an underwater environment}
In this section, we will derive the detection probability of Airy-OAM beam when it is passed through the underwater turbulent channel.

\begin{figure}[htbp]
\centerline{\includegraphics[width=1\columnwidth]{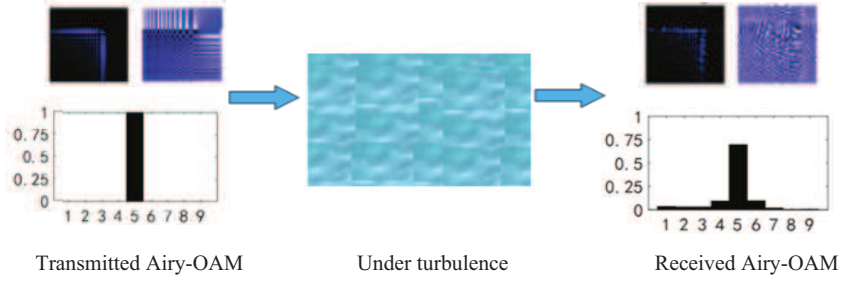}}
\caption{Schematic diagram of the propagation property of Airy-OAM beam in an underwater environment.}
\label{fig1}
\end{figure}

Fig.\ref{fig1} shows the schematic diagram of the propagation property of Airy-OAM beam in an underwater environment. At transmitter, Airy-OAM beam with topological charge $m_{0}$ was prepared by a special device, such as, spatial light modulator (SLM). Here, $m_{0}=5$ was used. The energy distribution at transmitter showed that all the energy was concentrated at $m_{0}=5$ position. Then, the Airy-OAM beam was passed through an underwater turbulence channel. Here, Rytov approximation model was adopted to describe the interference caused by the turbulent channel. The underwater turbulence would cause both phase and intensity fluctuations on Airy-OAM beam, resulting in the energy of Airy-OAM beam dispersed from $m=1$ to $m=9$. In order to estimate the mode dispersion caused by underwater turbulence, the detection probability $P_{m_{0}}$ is used to evaluate the property of Airy-OAM beam in the underwater environment, which is defined as
\begin{eqnarray}\label{EQ_1}
P_{m_{0}}= \frac{E_{m_{0}}}{\sum_{m} E_{m}},
\end{eqnarray}
where $E_{m_{0}}$ denotes the energy detected for the received Airy-OAM mode with $m_{0}$ topological charge, $m$ represents all possible received Airy-OAM modes.

The normalized complex amplitude of an Airy-OAM beam in the paraxial approximation can be expressed as \cite{guangshubiaodashi}
\begin{equation}
Ai_{0}(r,\varphi,z)=-\frac{ik}{z}\omega_{0}(r_{0}-\omega_{0}\alpha^{2})J_{m_{0}}(\frac{krr_{0}}{z})exp(ik\frac{r^{2}}{2z}+\frac{\alpha^{3}}{3}+im_{0}\varphi),
\label{Eq_1}
\end{equation}
where $(r, \varphi, z)$ are cylindrical coordinates, $r$ is a radial distance from the propagation axis,  $\varphi$ is an azimuthal angle, $z$ is a propagation distance.  $k=\frac{2\pi}{\lambda}$ is wavenumber and $\lambda$ is wavelength, $J_{m_{0}}(\cdot)$ is the Bessel function of the first kind. $\omega_{0}$ is associated with the arbitrary transverse scale, $r_{0}$ represents the radius of the main ring, $\alpha$ is the exponential truncation, $m_{0}$ represents topological charge, and it is OAM quantum number.

The influence caused by underwater turbulence can be regarded as pure interference on the phase \cite{lommel}. The second order cross spectral density function of the Airy-OAM beam with Rytov approximation can be expressed as
\begin{equation}
W(r,\varphi,\varphi^{'},z)=Ai_{0}(r,\varphi,z) \cdot Ai_{0}^{*}(r,\varphi^{'},z)exp[-\frac{1}{2}M(r,r^{'},z)],
\label{Eq_2}
\end{equation}
where $z$ is propagation distance, $M(r,r^{'},z)$ is the wave structure function. It could be represented as in \cite{lommel,bojiegou}
\begin{equation}
M(r,r^{'},z)=8\pi^{2}k^{2}z\int_{0}^{1}\int_{0}^{\infty}\kappa\Phi(\kappa,\xi)[1-J_{0}(\kappa|r-r^{'}|)]d\kappa d\xi=\frac{2|r-r^{'}|^{2}}{\rho_{c}^{2}},
\label{Eq_3}
\end{equation}
here, $J_{0}(\cdot)$ is the the first kind zero order Bessel function, $\Phi(\kappa,\xi)$ is the spectrum of anisotropic ocean turbulence, and the anisotropy is assumed only existing along the Airy-OAM beam's propagation direction, $\rho_{c}$ is spatial coherent radius.

In the deduction, $\Phi(\kappa,\xi)$ is adopted the expression in \cite{tuanliuyixing}, which is
\begin{equation}
\Phi(\kappa,\xi)=0.388\times10^{-8}\chi_{t}\zeta^{2}\varepsilon^{-1/3}\kappa^{-11/3}[1+2.35(\kappa\eta)^{2/3}]\times\phi(\kappa,\varpi),
\label{Eq_4}
\end{equation}
where $\kappa=\sqrt{\kappa_{z}^{2}+\zeta^{2}\kappa_{\rho}^{2}}$, $\kappa_{\rho}=\sqrt{\kappa_{x}^{2}+\kappa_{y}^{2}}$, $\kappa$ is the spatial frequency of turbulent fluctuation, $\chi_{t}$ is the dissipation rate of temperature variance ( the range is from $10^{-10}K^{2}s^{-1}$ to $10^{-4}K^{2}s^{-1}$) , $\zeta$ is the anisotropic coefficient, $\varepsilon$ is the rate of dissipation of kinetic energy per unit mass of fluid ( the range is from $10^{-10}m^{2}s^{-3}$ to $10^{-1}m^{2}s^{-3}$), $\eta$ is inner scale factor of oceanic turbulence, and $\phi(\kappa,\varpi)=[exp(-A_{T}\sigma)+\varpi^{-2}exp(-A_{S}\sigma)-2\varpi^{-1}exp(-A_{TS}\sigma]$, $A_{T}=1.863\times10^{-2}$, $A_{S}=1.9\times10^{-4}$, $A_{TS}=9.41\times10^{-3}$, $\sigma =8.284(\kappa\eta)^{4/3}+12.978(\kappa\eta)^{2}$, $\varpi$ is the ratio of temperature and salinity contributions to the refractive index spectrum (varying from $-5$ to $0$, corresponding to dominating temperature-induced or salinity-induced optical turbulence, respectively).

Substituting Eq.(\ref{Eq_4}) into Eq.(\ref{Eq_3}), we can derive the analytical expression of $\rho_{c}$  as
\begin{equation}
\rho_{c}^{2}=[8.705\times10^{-8}\kappa^{2}(\varepsilon\eta)^{-1/3}\zeta^{-2}\chi_{t}z(1-2.605\varpi^{-1}+7.007\varpi^{-2})]^{-1}.
\label{Eq_5}
\end{equation}

On the other hand, the second order cross spectral density function $W(r,\varphi,\varphi^{'},z)$ can be decomposed as
\begin{equation}
W(r,\varphi,\varphi^{'},z)=\frac{1}{2\pi}\sum_{-\infty}^{\infty}<|\varpi_{m}|^{2}>exp([im(\varphi-\varphi^{'})]),
\end{equation}
where $<|\varpi_{m}|^{2}>$  represents the ensemble averaged radial energy density of each harmonic component in oceanic turbulence.
Combining Eq.(7) with Eq.(2),Eq.(3),Eq.(4) and Eq.(6), and with the help of the integration formula
\begin{equation}
\int_{0}^{2\pi}exp[-in\varphi+\tau cos(\varphi-\varphi^{'})]d\varphi=2\pi exp(-in\varphi^{'})I_{n}(\tau),
\end{equation}
$<|\varpi_{m}|^{2}>$ can be obtained  as
\begin{eqnarray}
& &<|\varpi_{m}|^{2}>=\frac{1}{2\pi}\int_{0}^{2\pi}\int_{0}^{2\pi}W(r,\varphi,\varphi^{'},z)exp[-im(\varphi-\varphi^{'})]d\varphi d\varphi^{'} \nonumber\\
&=&\frac{1}{2\pi}\int_{0}^{2\pi}\int_{0}^{2\pi}|\frac{k}{z}\cdot \omega_{0}(r_{0}-\omega_{0}\alpha^{2})\cdot J_{m_{0}}(\frac{krr_{0}}{z})|^{2}\cdot exp(\frac{2\alpha^{3}}{3})exp[im_{0}(\varphi-\varphi^{'})] \nonumber\\
& &\cdot exp[-\frac{2r^{2}-2r^{2}cos(\varphi-\varphi^{'})}{\rho_{c}^{2}}]\cdot exp[-im(\varphi-\varphi^{'})]d\varphi d\varphi^{'}\nonumber\\
&=&\frac{2\pi k^{2}}{z^{2}}\omega_{0}^{2}(r_{0}-\omega_{0}\alpha^{2})^{2}|J_{m_{0}}(\frac{krr_{0}}{z})|^{2}exp(\frac{2\alpha^{3}}{3}-\frac{2r^{2}}{\rho_{c}^{2}})I_{m-m_{0}}(\frac{2r^{2}}{\rho_{c}^{2}}),
\end{eqnarray}
where $I_{m-m_{0}}(\cdot)$ is the symbol of the modified Bessel function of the first kind.

Assume the topological charge of transmitted and received Airy-OAM beam are $m_{0}$ and $m$ $(m\in(-\infty,\infty))$, respectively. With the integration of  $<|\varpi_{m}|^{2}>$ over the receiving aperture, we can obtain the energy of each received OAM mode as $E_{m}=\int_{0}^{R}<|\varpi_{m}|^{2}>rdr$, where $R$ is the size of the receiving aperture. Using the definition in Eq.(1), we can achieve the detection probability of transmitted Airy-OAM beam with topological charge $m_{0}$
when the Airy-OAM beam is propagated through the underwater turbulence channel as
\begin{eqnarray}
P_{m_{0}}&=&\frac{E_{m_{0}}}{\sum_{m}E_{m}}=\frac{\int_{0}^{R}<|\varpi_{m_{0}}|^{2}>rdr}{\sum_{m}\int_{0}^{R}<|\varpi_{m}|^{2}>rdr} \nonumber\\
&=&\frac{\int_{0}^{R}\frac{2\pi k^{2}}{z^{2}}\omega_{0}^{2}(r_{0}-\omega_{0}\alpha^{2})^{2}|J_{m_{0}}(\frac{krr_{0}}{z})|^{2}exp(\frac{2\alpha^{3}}{3}-\frac{2r^{2}}{\rho_{c}^{2}})I_{m_{0}-m_{0}}(\frac{2r^{2}}{\rho_{c}^{2}})rdr}{\sum_{m}\int_{0}^{R}\frac{2\pi k^{2}}{z^{2}}\omega_{0}^{2}(r_{0}-\omega_{0}\alpha^{2})^{2}|J_{m_{0}}(\frac{krr_{0}}{z})|^{2}exp(\frac{2\alpha^{3}}{3}-\frac{2r^{2}}{\rho_{c}^{2}})I_{m-m_{0}}(\frac{2r^{2}}{\rho_{c}^{2}})rdr}. \nonumber\\
\end{eqnarray}

\section{Numerical simulation and discussion}
\begin{figure}[!htbp]
\centerline{\includegraphics[width=0.8\columnwidth]{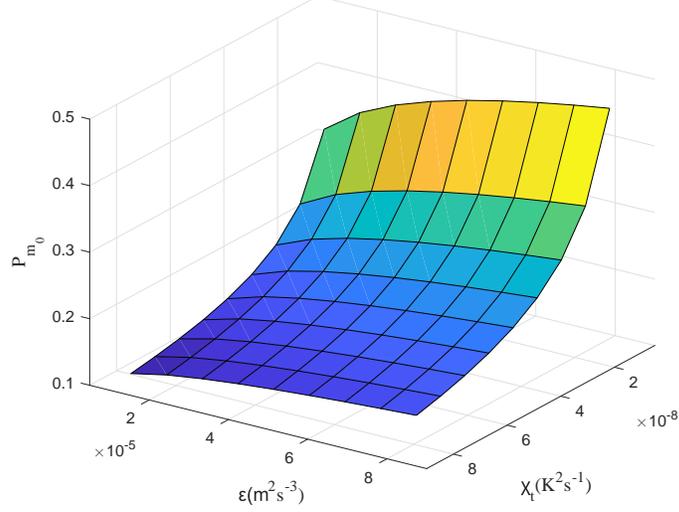}}
\caption{The detection probability $P_{m_{0}}$ under different dissipation rate of temperature variance $\chi_{t}$ and dissipation rate of kinetic energy per unit mass of fluid $\varepsilon$.}
\label{FIG2}
\end{figure}
In this section, the parameters of both underwater turbulence and Airy-OAM beam are explored to analyze the influence of underwater turbulence on the detection probability $P_{m_{0}}$.  Some parameters are set as the following in the numerical simulations. $\chi_{t}=10^{-8}K^{2}s^{-1}, \varepsilon=10^{-5}m^{2}s^{-3}, \eta=10^{-3}m, \zeta=2, \varpi=-3, \omega_{0}=10^{-2}m, r_{0}=10^{-3}m, \alpha=5\times10^{-2}, R=3\times10^{-2}m, m_{0}=1, \lambda=532\times10^{-9}m$ and $z=100m$.

We first analyze the influence of dissipation rate of temperature variance $\chi_{t}$ and dissipation rate of kinetic energy per unit mass of fluid $\varepsilon$ on the propagation of Airy-OAM beam. Fig.\ref{FIG2} shows the detection probability $P_{m_{0}}$ under different dissipation rate of temperature variance $\chi_{t}$ and dissipation rate of kinetic energy per unit mass of fluid $\varepsilon$. The results showed that with the increase of $\chi_{t}$ and the decrease of $\varepsilon$, the detection probability $P_{m_{0}}$ decreased. The detection probability $P_{m_{0}} $ was $0.4203$ and $0.179$, respectively, when $\chi_{t} $ was $10^{-8}K^{2}s^{-1}$ and $ 5\times10^{-8}K^{2}s^{-1}$, while $\varepsilon $ was the same, $5\times10^{-5}m^{2}s^{-3}$. The larger rate of dissipation of kinetic energy per unit mass of fluid $\varepsilon$ corresponded to lower oceanic turbulence.
It also indicated that the dissipation rate of temperature variance $\chi_{t}$  had more influence on the detection probability $P_{m_{0}}$ than that of dissipation rate of kinetic energy per unit mass of fluid $\varepsilon$.

\begin{figure}[htbp]
\centerline{\includegraphics[width=0.8\columnwidth]{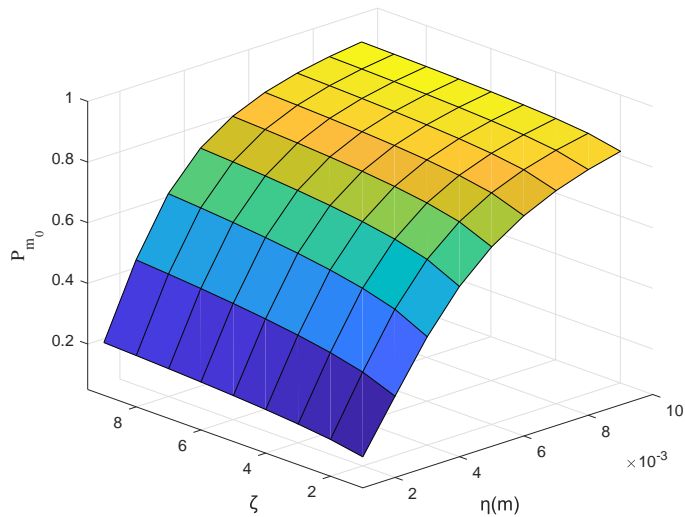}}
\caption{The detection probability $P_{m_{0}}$ under different inner scale factor $\eta$ and anisotropic coefficient $\zeta$ of oceanic turbulence.}
\label{FIG3}
\end{figure}
Later, we discuss the influence of inner scale factor $\eta$ and anisotropic coefficient $\zeta$ on the propagation of Airy-OAM beam. Fig.\ref{FIG3} shows the detection probability $P_{m_{0}}$ under different inner scale factor $\eta$ and anisotropic coefficient $\zeta$ of oceanic turbulence. Note that turbulence spectrum will degrade to the isotropic case when $\zeta$ is 1, and the turbulent heterogeneity will increase when $\zeta$ increases. The numerical results showed that $P_{m_{0}}$ increased with the increase of $\zeta$ for any given $\eta$. The reason was that the anisotropy of turbulence reduced refractive index fluctuation, so as to reduce the interference caused by oceanic turbulence. On the other hand, the detection probability $P_{m_{0}}$ became larger with the increase of $\eta$ when $\zeta$ was fixed. That was  the turbulence eddies in the inertial area would decrease with $\eta$, so that the beam scattering was reduced as the increase of $\eta$.

\begin{figure}[htbp]
\centerline{\includegraphics[width=0.8\columnwidth]{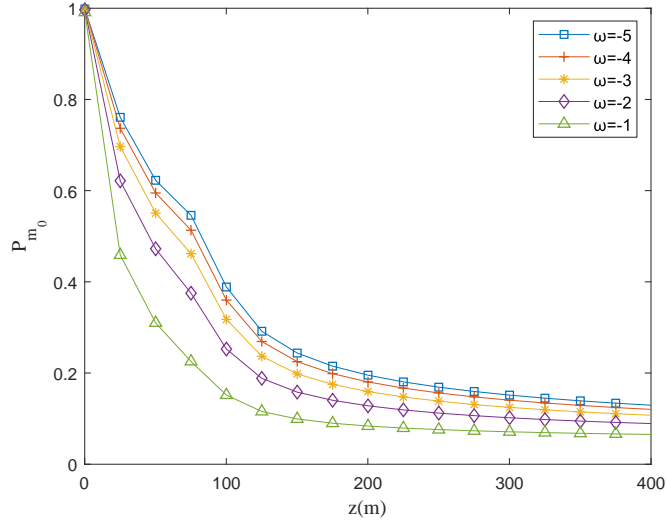}}
\caption{The detection probability $P_{m_{0}}$ against $z$ for different ratio of temperature and salinity contributions to the refractive index spectrum $\varpi$ of ocean turbulence.}
\label{FIG4}
\end{figure}
Simultaneously, we analyze the influence of oceanic turbulence temperature and salinity contributions ratio $\varpi$ on the propagation of Airy-OAM beam. Fig.\ref{FIG4} shows  the detection probability $P_{m_{0}}$ against $z$ for different ratio of temperature and salinity contributions to the refractive index spectrum $\varpi$. It was seen from the figure that $P_{m_{0}}$  decreased with the propagation distance $z$ for a given $\varpi$. For the same propagation distance, the detection probability $P_{m_{0}}$ was larger when $\varpi$ was smaller, which   implied that the salinity fluctuations in turbulence had more impact on the Airy-OAM beam than that of  temperature fluctuations. 

\begin{figure}[htbp]
\centerline{\includegraphics[width=0.8\columnwidth]{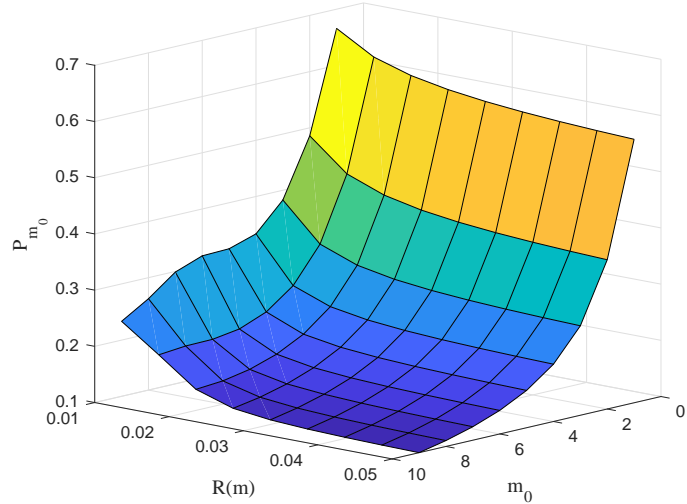}}
\caption{The detection probability $P_{m_{0}}$ under different topological charge $m_{0}$ and the size of the receiving aperture $R$.}
\label{FIG5}
\end{figure}
Furthermore, we demonstrate the influence of beam source parameters on the propagation of Airy-OAM beam. The influence of topological charge and the receiving aperture size on the  detection probability $P_{m_{0}}$ is shown in Fig.\ref{FIG5}. We could see that the detection probability $P_{m_{0}}$ decreased as topological charge $m_{0}$ and the receiving aperture size $R$ increased. The phenomenon could be explained that the radius of Airy-OAM beam increased with $m_{0}$, and the larger topological charge received Airy-OAM beam was limited by $R$. When the receiving aperture size $R$ was given, Airy-OAM beam with smaller topological charge had a higher detection probability. A larger receiving aperture size $R$ led to a smaller $P_{m_{0}}$.
\begin{figure}[htbp]
\centerline{\includegraphics[width=0.8\columnwidth]{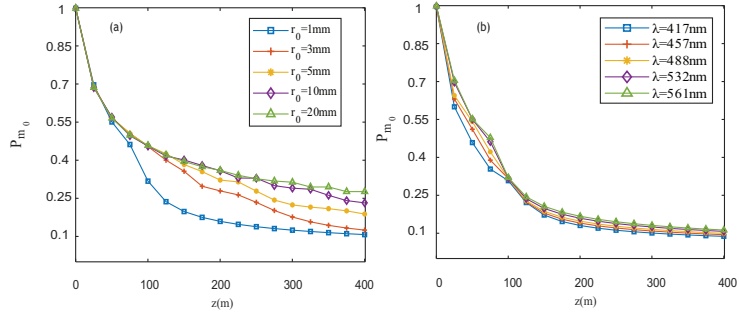}}
\caption{The detection probability $P_{m_{0}}$ against $z$ for different main ring radius $r_{0}$ (a) and different wavelength $\lambda$ (b) of Airy-OAM beam.}
\label{FIG6}
\end{figure}

Fig.\ref{FIG6} further shows the change of $P_{m_{0}}$ against the main ring radius $r_{0}$ and wavelength $\lambda$ when Airy-OAM beam is propagated through the oceanic turbulence channel, where Fig.\ref{FIG6}(a) is for different main ring radius, and Fig.\ref{FIG6}(b) is for different wavelengths. The result showed that the influence of $r_{0}$ on $P_{m_{0}}$ was not obvious when the propagation distance was small, say, less than $50m$. As the propagation distance increased, the dispersion degree of Airy-OAM beam decreased, resulting in the increase of the detection probability. It was also shown that the Airy-OAM beam with the main ring radius $r_{0}=20mm$ had the best performance, so that we set the parameter $r_{0}=20mm$ in the later comparisons. The results in Fig.\ref{FIG6}(b) also showed that the detection probabilities for different wavelengths were close, the detection probability of Airy-OAM beam with $\lambda=561nm$ was a little bigger than that of $\lambda=417nm$.

\begin{figure}[htbp]
\centerline{\includegraphics[width=0.8\columnwidth]{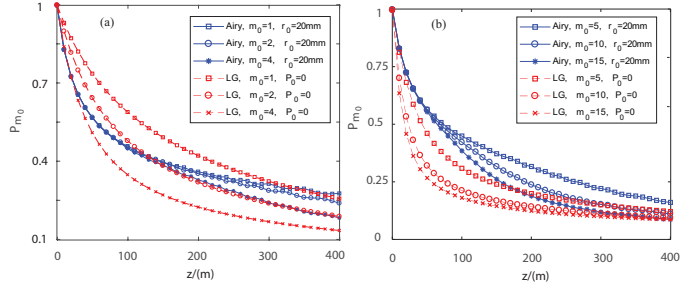}}
\caption{The detection probability $P_{m_{0}}$ against $z$ for Airy-OAM beam and LG beam under the same oceanic turbulence channel.}
\label{FIG7}
\end{figure}
At last, we compare the propagation property of Airy-OAM beam with those of Laguerre Gaussian(LG) beam under the same oceanic turbulence channel in Fig.\ref{FIG7}. The radial mode index $P_{0}$ of LG beam was $0$, and the radius of the main ring $r_{0}$ of Airy-OAM beam was $20mm$. The numerical results showed that the interference caused by oceanic turbulence on Airy-OAM beams was bigger than those on LG beams when topological charge $m_{0}$ is less than or equal to 4, which was consistent with those results in Ref.\cite{hermite}.
But the detection probability of Airy-OAM beam was larger than that of LG beam in Fig.\ref{FIG7}(b), when $m_{0}$ was equal or greater than $5$. The reason was Airy-OAM beam had non-diffraction and self-healing characteristics, its reconstruction ability would reduce the power loss caused by turbulence, and would increase the detection probability at the receiver side, when the topological charge of Airy-OAM beam was greater than $5$.

\section{Conclusion}

In this paper, we have demonstrated the propagation properties of Airy-OAM beam in an anisotropic weak oceanic turbulent channel. We have derived the analytic formula of detection probability for  Airy-OAM beam with Rytov approximation theory. The influences of turbulence parameters, beam parameters and propagation distance on the propagation property of Airy-OAM beam have been discussed. The results have shown that the interference caused by oceanic turbulence on the propagation of Airy-OAM beam become stronger as the dissipation rate of temperature variance, the ratio of temperature and salinity contributions to the refractive index spectrum, and the propagation distance increase. Simultaneously, the detection probability of Airy-OAM beam after oceanic turbulence have decreased as the dissipation rate of kinetic energy per unit mass of fluid, inner scale factor, and anisotropic coefficient decrease. Smaller topological charge, longer wavelength, and larger main ring radius Airy-OAM beams have more anti-interference to oceanic turbulence. Additionally, the detection probability of Airy-OAM beam after propagation has increased with the decreasing of receiving aperture size. In comparison with the LG-OAM beam, Airy-OAM beams have more resistance to turbulence when its topological charge is larger than 5. These results are useful for UOC link using Airy-OAM beam.

\section*{Acknowledgment}
The paper is supported by the National Natural Science Foundation of China (61475075,61271238), Postgraduate Research \& Practice Innovation Program of Jiangsu Province(KYCX18\underline{\hspace{0.5em}}0899).

\end{document}